\documentclass[lettersize,journal]{IEEEtran}
\usepackage{amsmath,amsfonts}
\usepackage{algorithmic}
\usepackage{algorithm}
\usepackage{array}
\usepackage[caption=false,font=normalsize,labelfont=sf,textfont=sf]{subfig}
\usepackage{textcomp}
\usepackage{stfloats}
\usepackage{url}
\usepackage{verbatim}
\usepackage{graphicx}
\usepackage{cite}
\usepackage{amssymb}
\usepackage{makecell}
\usepackage{balance}
\begin{document}

\title{Wireless-Powered Mobile Crowdsensing Enhanced by UAV-Mounted RIS: Joint Transmission, Compression, and Trajectory Design}
\author{Yongqing Xu,~\IEEEmembership{Graduated Student Member,~IEEE}, Haoqing Qi, Zhiqin Wang, Xiang Zhang, \\Yong Li,~\IEEEmembership{Member,~IEEE}, and Tony Q.S. Quek,~\IEEEmembership{Fellow,~IEEE}
\thanks{The work of Y. Xu, H. Qi, and Y. Li was supported by the National Research and Development Program ``Distributed Large Dimensional Wireless Cooperative Transmission Technology Research and System Verification'' under Grant 2022YFB2902400. The work of Tony Q. S. Quek was supported by the National Research Foundation, Singapore and Infocomm Media Development Authority under its Future Communications Research \& Development Programme. (Corresponding Author: Yong Li.)
	
	Y. Xu, H. Qi, and Y. Li are with the Key Laboratory of Universal Wireless Communications, Beijing University of Posts and Telecommunications, Beijing 100876, China. Email: \{xuyongqing; qihaoqing; liyong\}@bupt.edu.cn.
	
	Z. Wang and X. Zhang are with China Academy of Information and Communications Technology, Beijing 100191, China. Email: \{zhiqin.wang; zhangxiang1\}@caict.ac.cn.
	
	T. Q. S. Quek is with the Singapore University of Technology and Design, Singapore 487372 (e-mail: tonyquek@sutd.edu.sg).}}



\maketitle

\begin{abstract}
Mobile crowdsensing (MCS) enables data collection from massive devices to achieve a wide sensing range. Wireless power transfer (WPT) is a promising paradigm for prolonging the operation time of MCS systems by sustainably transferring power to distributed devices. However, the efficiency of WPT significantly deteriorates when the channel conditions are poor. Unmanned aerial vehicles (UAVs) and reconfigurable intelligent surfaces (RISs) can serve as active or passive relays to enhance the efficiency of WPT in unfavourable propagation environments. Therefore, to explore the potential of jointly deploying UAVs and RISs to enhance transmission efficiency, we propose a novel transmission framework for the WPT-assisted MCS systems, which is enhanced by a UAV-mounted RIS. Subsequently, under different compression schemes, two optimization problems are formulated to maximize the weighted sum of the data uploaded by the user equipments (UEs) by jointly designing the WPT and uploading time, the beamforming matrics, the CPU cycles, and the UAV trajectory. A block coordinate descent (BCD) algorithm based on the closed-form beamforming designs and the successive convex approximation (SCA) algorithm is proposed to solve the formulated problems. Furthermore, to highlight the insight of the gains brought by the compression schemes, we analyze the energy efficiencies of compression schemes and confirm that the gains gradually reduce with the increasing power used for compression. Simulation results demonstrate that the amount of collected data can be effectively increased in wireless-powered MCS systems.
\end{abstract}


\section{Introduction}
\IEEEPARstart{W}{ith} the development of wireless networks, numerous application scenarios have emerged, such as smart cities, smart homes, smart transportation, smart grids, and smart healthcare \cite{wang2024crowdsensing}. By connecting and organizing numerous terminals within these scenarios, Internet-of-Things (IoT), ambient IoT (A-IoT), and artificial intelligence of things (AIoT) networks are established, where terminals have the abilities to sense, compute, and exchange information to make intelligent decisions \cite{butt2024ambient,zhang2020empowering}. Mobile crowdsensing (MCS) is a crucial technology for IoT, A-IoT, and AIoT, enabling mobile phones and wearable smart devices to upload sensing data, aggregating and fusing the sensed data in the cloud to achieve comprehensive area sensing and intelligent decision-making \cite{capponi2019survey}. Research on MCS is currently underway to investigate network architectures, resource allocation schemes, and frame structures of MCS systems. For example, the authors of \cite{li2022joint} proposed a joint sensing, communication, and computation framework for MCS systems, where the transmission parameters and CPU cycles are jointly optimized. The authors of \cite{zhou2022joint} considered an energy-efficient transmission scheme, which can minimize the energy consumption for data sensing and transmission while satisfying the minimal sensing and transmission requirements. Despite the numerous benefits of MCS, the energy constraints of many devices with a small or no battery result in short and unstable MCS operation times.

\subsection{Wireless-Powered Mobile Crowdsensing}
Wireless power transfer (WPT) is an emerging paradigm that enables delivering energy wirelessly \cite{clerckx2021wireless,guo2018wireless}. WPT can prolong the operational time of MCS systems by charging distributed devices via electromagnetic waves \cite{wang2016adaptively}. The core of WPT-powered MCS lies in using energy to exchange sensing information, thereby motivating more device involvement. Specifically, the authors of \cite{galinina2018wirelessly} developed an energy-data trading mechanism for WPT-powered MCS systems and offered a corresponding network architecture. In \cite{li2018wirelessly}, the authors jointly controlled the wireless-power allocation at the base station (BS), the sensing and compression parameters of each user equipment (UE) to maximize data utility and minimize energy consumption simultaneously. Moreover, in \cite{le2023wirelessly}, a WPT-powered MCS-assisted sustainable federated learning architecture is proposed, where an optimization model is established to minimize the total task completion time. However, the efficiency of WPT depends on the hardware circuitry and channel conditions, and it significantly deteriorates in scenarios with poor channel conditions, such as emergency rescue, traffic inspection, and forest protection scenarios \cite{whitepaper}.

\subsection{UAV-Assisted Wireless-Powered Mobile Crowdsensing}
Unmanned aerial vehicle (UAV) is recognized as a promising technology for WPT systems because the efficiency of WPT can be significantly enhanced by the high mobility, flexibility, and high probability of the line-of-sight (LoS) path associated with UAVs \cite{qi2024uav,xu2018uav,yuan2021joint,li2023energy}. Specifically, the authors of \cite{xu2018uav} derived optimal UAV locations to maximize the sum of harvested energy. The authors of \cite{yuan2021joint} jointly designed the UAV trajectory and the orientation of the UAV antenna to maximize the harvested energy. Moreover, in \cite{li2023energy}, the authors modelled the UAV's power consumption and proposed two types of UAV trajectories, which can maximize the power transferred by the UAV and guarantee that the UAV safely flies to the BS. More importantly, MCS systems can also benefit from the advantages of UAVs. For example, a joint WPT, compression, and data uploading framework for UAV-assisted MCS systems was proposed in \cite{qi2024uav}. In \cite{zhou2018computation}, the authors formulated an uploading rate maximization problem in a UAV-enabled WPT system and proposed an alternative algorithm to solve the formulated non-convex problem. In \cite{he2023energy}, the authors considered a two-stage transmission framework for a UAV-enabled WPT system, where the time durations of two stages and transmit parameters are optimized. Furthermore, the authors of \cite{liu2021uav} considered a more practical UAV-assisted MCS scenario, where the outage performance of the WPT and data rate are analyzed, and an outage probability minimization problem under specific constraints was formulated and solved. All the aforementioned works assume that UAVs have sufficient energy to perform WPT and data reception. However, UAVs are powered by onboard batteries, and the manoeuvring actions consume most of the UAVs' energy \cite{8486942}, which limits the UAV capacity for WPT and data reception.

\subsection{RIS and UAV-Assisted Wireless-Powered Mobile Crowdsensing}
Reconfigurable intelligent surface (RIS) is emerging as a critical technology for next-generation wireless networks due to its cost-effective and energy-effective characteristics \cite{tran2022multifocus,yang2021reconfigurable,xu2023joint,xu2024robust}. By integrating UAV with RIS, preferable radio environments for WPT systems can be created \cite{zhao2020wireless,tran2022reconfigurable,cheng2022self}. The efficiency of WPT has the potential to be further enhanced \cite{pang2021uav}, e.g., the WPT protocols for RIS-assisted UAV systems proposed by the authors of \cite{ren2022energy} and \cite{peng2023energy} can enhance the WPT efficiency comparing with the schemes without RIS. Furthermore, the integration of RIS and UAV in MCS systems has garnered significant interest because of the high transmission efficiency \cite{zhai2023ris,liu2021joint,chen2023ris,nguyen2022ris,zhou2023flying,liu2022flexible,lin2024ergodic}. Research on RIS and UAV-assisted MCS can be categorized into two types. The first type is the MCS assisted by separate deployment of RISs and UAVs, such as \cite{zhai2023ris,liu2021joint,chen2023ris,nguyen2022ris}, where RISs are installed on the wall of tall buildings. The authors focused on the joint optimization of UAV transmit power, UAV trajectory, UE time allocation, and RIS beamforming matrix. The second type is the MCS assisted by UAV-mounted RISs, such as \cite{zhou2023flying,liu2022flexible,lin2024ergodic}, where RISs are mounted on UAVs and the authors shifted their research focus to the joint design of the transmission parameters of BSs and UEs, the beamforming matrics of RISs, and the UAV trajectories. The design complexity of MCS systems assisted by separate deployment of RIS and UAV is low. However, the onboard batteries of UAVs limit their application as active aerial relays. In UAV-mounted RIS-assisted MCS systems, the UAVs act as passive aerial relays, which can enhance transmission efficiency. Research on UAV-mounted RIS-assisted MCS systems is still in its infancy. There remains significant potential to further explore and harness the capabilities of UAV-mounted RIS-assisted MCS technology.

\subsection{The Contributions of This Work}
Motivated by the above-mentioned works, we first introduce the computational capabilities of UEs into UAV-mounted RIS-assisted MCS systems and establish a transmission framework for wireless-powered MCS systems to maximize the amount of data uploaded by UEs. The main contributions of  our work are listed as follows:
\begin{itemize}
	\item We design a transmission model based on the developed time-division multiple access (TDMA) protocol for UAV-mounted RIS-assisted MCS systems. We also establish the WPT, compression, and communication models for the considered wireless-powered MCS system enhanced by a UAV-mounted RIS.
	\item We establish two optimization problems to maximize the amount of uploading data under two compression schemes. Both problems require to jointly optimize the WPT time of the BS, the transmit and receive beamforming vectors of the BS, the RIS beamforming matrices, the CPU cycles of UEs, the transmit power of UEs, the uploading times of UEs, and the UAV trajectory. Furthermore, we propose a block coordinate descent (BCD) algorithm to solve the formulated problems, which can iteratively converge to a near-optimal solution.
	\item To highlight the gains brought by lossless compression and the performance losses caused by lossy compression, we analyze the energy efficiencies of both methods by comparing them with the direct uploading scheme. Moreover, we analyze the computational complexity for solving the formulated optimization problems and the convergence performance of the BCD algorithm.
	\item We also perform numerous simulations from different perspectives to verify the effectiveness of the proposed transmission model and algorithms.
\end{itemize}

\subsection{Paper Organization}
The remainder of this paper is organized as follows. Section \uppercase\expandafter{\romannumeral2} introduces the system model and performance metrics. Section \uppercase\expandafter{\romannumeral3} proposes a BCD-based algorithm to design the parameters of BS, RIS, UEs, and UAV under the lossless compression model. Section \uppercase\expandafter{\romannumeral4} formulates a problem for maximizing the amount of uploading data under the lossy compression model. Section \uppercase\expandafter{\romannumeral5} analyzes the energy efficiencies. Section \uppercase\expandafter{\romannumeral6} analyzes the complexity and convergence performance of the proposed algorithms. Section \uppercase\expandafter{\romannumeral7} presents the simulation results. Section \uppercase\expandafter{\romannumeral8} concludes the paper.

\subsection{Notations}
Throughout this paper, matrices are denoted by bold uppercase letters, e.g., $\boldsymbol{X}$, vectors are denoted by bold lowercase letters, e.g., $\boldsymbol{x}$, and scalars are denoted by normal fonts, e.g., $x$. $\left\vert\cdot\right\vert$ represents the absolute value operation. $\left\Vert\cdot\right\Vert$ represents Euclidean norm of a vector. $\left(\cdot\right)^T$, $\left(\cdot\right)^H$, and $\left(\cdot\right)^*$ represent the transpose, the conjugate transpose, and the conjugate, respectively. $\mathbb{C}^{\left(\cdot\right)}$ denotes the complex space. $\mathbb{R}^{\left(\cdot\right)}$ denotes the real space. $\text{diag}\left(\boldsymbol{x}\right)$ represents using $\boldsymbol{x}$ to formulate a diagonal matrix. $\text{arg}\left(\cdot\right)$ represents the angle of a complex number.

\section{System Model}
\begin{figure}[!t]
	\centering
	\includegraphics[width=0.5\textwidth]{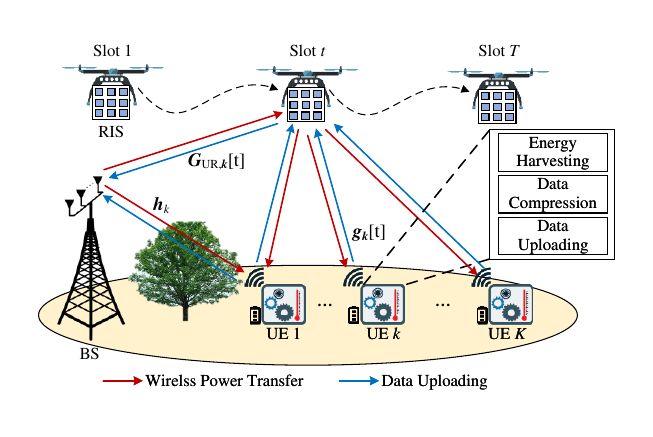}
	\caption{Considered wireless-powered MCS system enhanced by a UAV-mounted RIS.}
	\label{fig1}
\end{figure}
The considered scenario is shown in Fig. 1, where a BS equipping $M$ antennas transfers power to $K$ single-antenna UEs and receives data from the UEs. A UAV mounting a $N$-element RIS assists the power transfer of BS and the data uploading of UEs during its flying duration (i.e., $T$ time slots). The BS and the RIS antenna arrays are configured as uniform linear arrays (ULAs). The inter-distance of the BS antennas and the RIS elements are half-wavelength. Under a three-dimensional (3-D) Cartesian coordinate system, the coordinates of the BS and the $k$-th UE are denoted as $\boldsymbol{q}_0=\left[x_{\text{B}},y_{\text{B}},z_{\text{B}}\right]^T\in\mathbb{R}^{3\times1}$ and $\boldsymbol{q}_k=\left[x_{k},y_{k},z_{k}\right]^T\in\mathbb{R}^{3\times1}$, respectively. The coordinate of the UAV at the $t$-th slot is $\boldsymbol{q}[t]=\left[x[t],y[t],z\right]^T\in\mathbb{R}^{3\times1}$. The heights of the BS, the UEs, and the UAV are all assumed to remain fixed during the flying duration of the UAV. The sets of UEs and time slots are represented by $\cal K$ and $\cal T$.

\begin{figure}[!t]
	\centering
	\includegraphics[width=0.5\textwidth]{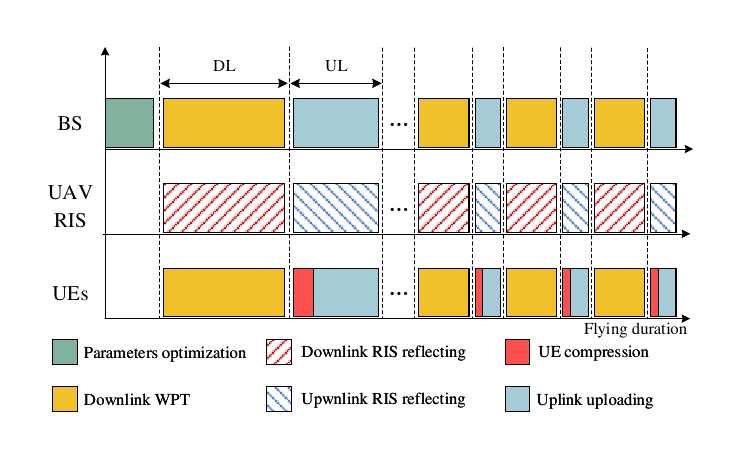}
	\caption{The proposed TDMA protocol for the considered UAV-mounted RIS-enhanced wireless-powered MCS system.}
	\label{fig_TDMA}
\end{figure}

The UEs harvest the transferred energy by the BS via their energy harvesting modules. Then, the UEs compress their collected sensing data and upload the data to the BS using the harvested energy via the data compression modules and the data uploading modules\footnote{For energy efficiency, only a part of the sensing data is compressed by the UEs. The details are discussed in Section \uppercase\expandafter{\romannumeral5}.}. To avoid self-interference, the wireless power transfer of the BS and the data uploading of the UEs are scheduled in a time-division multiplexing (TDM) manner. The proposed TDMA protocol is shown in Fig. \ref{fig_TDMA}.

\subsection{Wireless Power Transfer Model}
The harvested energy of the $k$-th UE at the $t$-th time slot can be expressed as
\begin{equation}
	\label{eq1}
	{E_k}[t] = {\eta _0}{\delta _t}{\left| {\left( {\boldsymbol{h}_k^H + \boldsymbol{g}_k^H[t]{\boldsymbol{\Theta}_k}[t]\boldsymbol{G}_{{\rm{UR}}}^H[t]} \right){\boldsymbol{w}_k}[t]} \right|^2}{a_k}[t],\forall k\in\mathcal{K},
\end{equation}
where $\eta_0$ denotes the energy conservation efficiency; $\delta_t$ denotes the length of one time slot; $\boldsymbol{h}_k=h_k\boldsymbol{a}\left(\phi_k\right)\in\mathbb{C}^{M\times1}$ denotes the channel vector between the BS and the $k$-th UE; $\boldsymbol{g}_k=g_k\boldsymbol{b}\left(\varphi_k\right)\in\mathbb{C}^{N\times1}$ and  $\boldsymbol{G}_{\text{UR}}[t]=G_0\boldsymbol{a}\left(\phi_0[t]\right)\boldsymbol{b}^T\left(\varphi_0[t]\right)\in\mathbb{C}^{M\times N}$ denote the channels between the RIS and the $k$-th UE and between the BS and the RIS at the $t$-th time slot; $h_k=\sqrt{\frac{h_0}{d_{\text{B2U,k}}^{\alpha_{\text{BU}}}}}$, $g_k[t]=\sqrt{\frac{h_0}{d_{\text{R2U},k}^{\alpha_{\text{RU}}}[t]}}$, and $G_0[t]=\sqrt{\frac{h_0}{d_{\text{B2R}}^{\alpha_{\text{BR}}}[t]}}$ denote the large scale channel gains, where $h_0$ represents the channel gain at the reference distance of one meter,  $\alpha_{\text{BU}}$, $\alpha_{\text{RU}}$, and $\alpha_{\text{BR}}$ are the pathloss exponents, $d_{\text{B2U,k}}=\left\Vert\boldsymbol{q}_k-\boldsymbol{q}_0\right\Vert$, $d_{\text{R2U},k}[t]=\left\Vert\boldsymbol{q}[t]-\boldsymbol{q}_k\right\Vert$, and $d_{\text{B2R}}[t]=\left\Vert\boldsymbol{q}[t]-\boldsymbol{q}_0\right\Vert$; $\boldsymbol{a}\left(\phi\right)=\left[1,\cdots,e^{-j\pi\left(N-1\right)\sin(\phi)}\right]^T\in\mathbb{C}^{M\times1}$ and $\boldsymbol{b}\left(\varphi\right)=\left[1,\cdots,e^{-j\pi\left(M-1\right)\sin(\varphi)}\right]^T\in\mathbb{C}^{N\times1}$ denote the array steering vectors of the BS antenna array and the RIS; $\phi_k$ and $\varphi_k[t]$ represent the angles of departure (AODs) of the signals transmitted and reflected by the BS to the $k$-th UE and by the RIS to the $k$-th UE at the $t$-th time slot; $\phi_0[t]$ and $\varphi_0[t]$ denote the AODs between the BS and the RIS and between the RIS and the BS at the $t$-th time slot; the sine of the AODs can be written as $\sin(\phi_k)=\frac{x_k}{d_{\text{B2U},k}}$, $\sin(\varphi_k[t])=\frac{\sqrt{d_{\text{R2U},k}^2[t]-(z-z_k)^2}}{d_{\text{R2U},k}[t]}$, $\sin\left(\phi_0[t]\right)=\frac{z-z_{\text{B}}}{d_{\text{B2R}}[t]}$, $\sin(\varphi_0[t])=\frac{\sqrt{d_{\text{B2R}}^2[t]-(z-z_{\text{B}})^2}}{d_{\text{B2R}}[t]}$. Moreover, $\boldsymbol{w}_k[t]\in\mathbb{C}^{M\times1}$ denotes the transmitting beamforming vector of the $k$-th UE at the $t$-th time slot; $\boldsymbol{\Theta}_k[t]=\text{diag}\left(\boldsymbol{\theta}_k[t]\right)\in\mathbb{C}^{N\times N}$ denotes the RIS beamforming matrix used to serve the WPT for the $k$-th UE at the $t$-th time slot; $0\leq a_k[t]\leq1$ denotes the time ratio of the $t$-th time slot occupied by the $k$-th UE for harvesting energy.

\subsection{Compression Model}
In this subsection, we introduce two types of the compression model. The first one is the lossless compression model, which aims to evaluate the performance upper bound obtained by the compression scheme. The second one is the lossy compression model, which aims to investigate the performance loss caused by imperfect compression.

\subsubsection{Lossless compression model}
Lossless compression refers to the fact that the compressed data can perfectly reconstruct the raw sensing data. Many compression methods, such as Huffman, run-length, and Lempel-Ziv encoding, can achieve lossless compression. We assume that all UEs adopt the same compression method and that their compression ratios are equal. The energy used for compression of the $k$-th UE at the $t$-th time slot is defined as \cite{liu2022flexible}
\begin{equation}
	\label{eq2}
	{E_{{\rm{C,}}k}}[t] \triangleq \gamma {\delta _t}{\gamma _c}f_k^3[t],\forall k\in\mathcal{K},
\end{equation}
where $\gamma$ denotes the time ratio of the compression at each time slot, $\gamma _c$ denotes a constant depending on the hardware, $f_k[t]$ denotes the CPU-cycle frequency of the $k$-th UE at the $t$-th slot. The total sensing data being compressed is defined as
\begin{equation}
	\label{eq3}
	{S_k}[t] \triangleq \frac{{\gamma {\delta _t}{f_k}[t]}}{{{e^{\varepsilon /{\kappa _k}}} - {e^\varepsilon }}},\forall k\in\mathcal{K},
\end{equation}
where $e^{\varepsilon /{\kappa _k}}-e^\varepsilon$ denotes the total CPU cycles that needs to compress one bit sensing data with $\varepsilon$ denoting a constant depending on the compression method, $\kappa_k\in\left(0,1\right]$ denotes the ratio of the compressed data to the raw data before compression. It is noted that the compression energy consumption is proportional to the cube of the compressed data volume. Therefore, the compression scheme can increase the energy efficiency when the compressed data volume is small. On the contrary, it decreases the energy efficiency when the compressed data volume is relatively large. We will quantitatively analyze the energy efficiency in  Section \uppercase\expandafter{\romannumeral5}. 

\subsubsection{Lossy compression model}
Lossy compression refers to the fact that the compressed data imperfectly reconstruct the raw sensing data due to the imperfections of hardware and compression algorithms. We first define the information loss caused by the lossy compression as \cite{li2018wirelessly}
\begin{equation}
	\label{eq4}
	l_k\triangleq1-\sqrt{\bar{\kappa}_k},\forall k\in\mathcal{K},
\end{equation}
where $\bar{\kappa}_k\in\left(0,1\right]$ denotes the ratio of the compressed data to the raw data before the lossy compression. Specifically, $l_k=0$ when $\bar{\kappa}_k=1$, which represents no sensing data has been compressed. $l_k\rightarrow1$ when $\bar{\kappa}_k\rightarrow0$, which represents all sensing data has been lost due to the compression. Then, the sensing data can be effectively compressed is expressed as
\begin{equation}
	\label{eq5}
	{\bar{S}_k}[t] \triangleq \frac{{\gamma {\delta _t}{\bar{f}_k}[t]}\sqrt{\bar{\kappa}_k}}{{{e^{\varepsilon /{\bar{\kappa} _k}}} - {e^\varepsilon }}},\forall k\in\mathcal{K},
\end{equation}
where $\bar{f}_k[t]$ denotes the CPU-cycle frequency under the lossy compression scheme.

\subsection{Communication Model}
The data uploaded by the $k$-th UE at the $t$-th slot is
\begin{equation}
	\label{eq6}
	\begin{aligned}
		&{R_k}[t] = B{\delta _t}{b_k}[t]\\
		&\cdot\!{\log _2}\!\left( {\!1\!\! + \!\!\frac{{{{\left| {\left( {\boldsymbol{h}_k^H + \boldsymbol{g}_k^H[t]{{\bar {\boldsymbol{\Theta}} }_k}[t]\boldsymbol{G}_{{\rm{UR}}}^H[t]} \right){\boldsymbol{u}_k}[t]} \right|}^2}{P_k}[t]}}{{\sigma _\text{B}^2}}}\! \right)\!,\!\forall k\!\in\!\mathcal{K},
	\end{aligned}
\end{equation}
where $B$ denotes the bandwidth of uplink; $0\leq b_k[t]\leq1$ denotes the time ratio occupied by the $k$-th UE at the $t$-th time slot for uploading data; $\bar{\boldsymbol{\Theta}}_k[t]=\text{diag}\left(\bar{\boldsymbol{\theta}}_k[t]\right)\in\mathbb{C}^{N\times N}$ denotes the RIS beamforming matrix used to serve the uploading of the $k$-th UE at the $t$-th slot; $P_k[t]$ denotes the transmit power of the $k$-th UE at the $t$-th slot; $\sigma _\text{B}^2$ is the noise variance at the BS; $\boldsymbol{u}_k[t]\in\mathbb{C}^{M\times1}$ denotes the receive beamforming vector of the $k$-th UE.

\noindent\textbf{\emph{\underline{Remark }1: }}\emph{The uploading data consists of the compressed data and the data without compression. Specifically, the amount of the raw sensing data uploaded by the $k$-th UE under the lossless compression scheme is $R_k[t]+(1-\kappa_k)S_k[t]$, where $(1-\kappa_k)S_k[t]$ represents the amount of the raw sensing data saved due to compression. The amount of the raw sensing data uploaded by the $k$-th UE under the lossy compression scheme is $R_k[t]+(1-\bar\kappa_k)\bar S_k[t]$.}

Our goal is to maximize the amount of the raw uploading data of all UEs by joint designing the BS transmit and receive beamforming vectors, the RIS beamforming matrics, the UAV trajectory, the CPU cycles, transmit power, energy harvesting time, and uploading time of all UEs.

\section{Uploading Data Maximization Based on Lossless Compression}
\subsection{Problem Formulation}
The problem of maximizing the amount of the total uploading data based on lossless compression can be formulated as
\begin{subequations}
	\label{eq7}
	\begin{align}
		&\text{(P1)}\mathop{\text{max}}\limits_{\boldsymbol{X},\boldsymbol{Y},\boldsymbol{Q}}\ \sum\limits_{t = 1}^T {\sum\limits_{k = 1}^K {{\lambda_k}} \left\{ {{R_k}[t] + \left( {1 - {\kappa _k}} \right){S_k}[t]} \right\}} \label{eq7a} \tag{7a} \\
		&\qquad\ \ \text{s.t.}\quad{f_k}[t],{P_k}[t],{a_k}[t],{b_k}[t] \ge 0,\forall k\! \in\! {\cal K},\forall t \!\in\! {\cal T},\label{eq7b} \tag{7b} \\
		&\qquad\qquad\ \ \left\| {{\boldsymbol{w}_k}[t]} \right\|^2 \le {P_{\rm{T}}},\forall t \in {\cal T},\forall k \in {\cal K},\label{eq7c} \tag{7c}\\
		&\qquad\qquad\ \ \left| {{\theta _{k,n,n}}[t]} \right| = 1,\forall t \in {\cal T},\forall k \in {\cal K},\label{eq7d} \tag{7d}\\
		&\qquad\qquad\ \ \sum\limits_{i = 1}^t {\left[ {\gamma {\delta _t}{\gamma _c}f_k^3[i] + {b_k}[i]{\delta _t}{P_k}[i]} \right]} \nonumber\\
		&\qquad\qquad\qquad\le \sum\limits_{i = 1}^t{E_k[i]},\forall k \in {\cal K},\forall t \in {\cal T},\label{eq7e} \tag{7e}\\
		&\qquad\qquad\ \ \sum\limits_{i = 1}^t {\frac{{\gamma {\delta _t}{f_k}[i]{\kappa _k}}}{{{e^{\varepsilon /{\kappa _k}}} - {e^\varepsilon }}}}  \!\le\! \sum\limits_{i = 1}^t {{R_k}[i]},\!\forall k \!\in\! {\cal K},\!\forall t \!\in\! {\cal T}\!,\!\label{eq7f} \tag{7f}\\
		&\qquad\qquad\ \ \sum\limits_{k = 1}^K {{a_k}[t]}  + \sum\limits_{k = 1}^K {{b_k}[t]}  \le 1,\forall t \in {\cal T},\label{eq7g} \tag{7g}\\
		&\qquad\qquad\ \ \sum\limits_{k = 1}^K {{b_k}[t]}  \le 1 - \gamma ,\forall t \in {\cal T},\label{eq7h} \tag{7h}\\
		&\qquad\qquad\ \ \left\| {\boldsymbol{q}\left[ {t + 1} \right] - \boldsymbol{q}\left[ t \right]} \right\| \le {V_{\max }}{\delta _t},\forall t \in {\cal T},\label{eq7i} \tag{7i}\\
		&\qquad\qquad\ \ \boldsymbol{q}\left[ 1 \right] = {\boldsymbol{q}_1},\boldsymbol{q}\left[ T \right] = {\boldsymbol{q}_T}\label{eq7j} \tag{7j},
	\end{align}
\end{subequations}
where $\boldsymbol{X}\triangleq\left\{ {{f_k}[t],{P_k}[t],{a_k}[t],{b_k}[t],\forall k \in {\cal K},\forall t \in {\cal T}} \right\}$, $\boldsymbol{Y}\triangleq\left\{ {{\boldsymbol{w}_k}[t],{\boldsymbol{u}_k}[t],{\boldsymbol{\Theta}_k}[t],{{\bar {\boldsymbol{\Theta}} }_k}[t],\forall k \in {\cal K},t \in {\cal T}} \right\}$, $\boldsymbol{Q}\triangleq\left\{ {\boldsymbol{q}\left[ t \right],\forall t \in {\cal T}} \right\}$, and $\lambda_k$ is the weight of the $k$-th UE. The objective function in (\ref{eq7a}) aimes to maximize the weighted sum of the amount of the raw uploading data of all UEs. The constraints in (\ref{eq7c}) guarantee that the transmit power of the BS at each time slot is less than the transmit power budget $P_{\text{T}}$. The constraints in (\ref{eq7d}) denote that the reflection magnitude of each RIS element at each time slot should be less than one. The constraints in (\ref{eq7e}) guarantee that the accumulated energy at each UE is more than its consumed energy for compressing and uploading. The constraints in (\ref{eq7f}) constrain the amount of compressed data of each UE to be less than its uploading rate. Moreover, to guarantee the WPT, the uploading, and the compression are scheduled in a TDM manner, we introduce the constraints in (\ref{eq7g}) and (\ref{eq7h}). The constraints in (\ref{eq7g}) guarantee that the WPT and uploading of all UEs are scheduled in a TDM manner. The constraints in (\ref{eq7h}) guarantee the uploading and compression of all UEs are scheduled in a TDM manner. The constraints in (\ref{eq7i}) constrain the UAV speed is less than $V_{\text{max}}$. The two constraints in (\ref{eq7j}) constrain the UAV initial and final positions. Due to the coupled variables and the non-convex constraints, (P1) is a non-convex problem. A BCD algorithm is developed to facilitate solving (P1).

\subsection{Subproblem of Designing CPU-cycle, Uploading Power, and Time Allocation}
To design $\boldsymbol{X}$, we first keep $\boldsymbol{Y}$ and $\boldsymbol{Q}$ fixed in this subsection. The problem of designing $\boldsymbol{X}$ can be reformulated as
\begin{subequations}
	\label{eq8}
	\begin{align}
		&\text{(P2)}\mathop{\text{max}}\limits_{\boldsymbol{X}}\ \sum\limits_{t = 1}^T {\sum\limits_{k = 1}^K {{\lambda_k}} \left\{ {{R_k}[t] + \left( {1 - {\kappa _k}} \right){S_k}[t]} \right\}} \label{eq8a} \tag{8a} \\
		&\qquad\ \text{s.t.}\quad \text{(\ref{eq7b}), (\ref{eq7e}), (\ref{eq7f}), (\ref{eq7g}), (\ref{eq7h})}.\label{eq8b} \tag{8b}
	\end{align}
\end{subequations}
It is observed that (P2) is still non-convex due to the coupled $b_k[t]$ and $P_k[i]$. We define $p_k[t]\triangleq b_k[t]P_k[t]$. Then, the uploading data of the $k$-th UE can be rewritten as
\begin{equation}
	\label{eq9}
	\begin{aligned}
		&{\bar R_k}[t] = B{\delta _t}{b_k}[t]\\
		&\cdot\!{\log _2}\!\left( {\!1\!\! + \!\!\frac{{{{\left| {\left( {\boldsymbol{h}_k^H + \boldsymbol{g}_k^H[t]{{\bar {\boldsymbol{\Theta}} }_k}[t]\boldsymbol{G}_{{\rm{UR}}}^H[t]} \right){\boldsymbol{u}_k}[t]} \right|}^2}{p_k}[t]}}{{b_k[t]\sigma _\text{B}^2}}}\! \right)\!,\!\forall k\!\in\!\mathcal{K}.
	\end{aligned}
\end{equation}
(P2) can be formulated as
\begin{subequations}
	\label{eq10}
	\begin{align}
		&\text{(P3)}\mathop{\text{max}}\limits_{\boldsymbol{X}/P_k[t],\atop\left\{p_k[t]\right\}}\ \sum\limits_{t = 1}^T \sum\limits_{k = 1}^K {{\lambda_k}} \left\{{\bar R_k}[t]\!+\! \left( {1 \!-\! {\kappa _k}} \right){S_k}[t] \right\} \label{eq10a} \tag{10a} \\
		&\qquad\ \text{s.t.}\quad{f_k}[t],{p_k}[t],{a_k}[t],{b_k}[t] \ge 0,\forall k\! \in\! {\cal K},\forall t \!\in\! {\cal T},\label{eq10b} \tag{10b} \\
		&\qquad\qquad\ \sum\limits_{i = 1}^t {\left[ {\gamma {\delta _t}{\gamma _c}f_k^3[i] + {\delta _t}{p_k}[i]} \right]} \nonumber\\
		&\qquad\qquad\qquad\le \sum\limits_{i = 1}^t{E_k[i]},\forall k \in {\cal K},\forall t \in {\cal T},\label{eq10c} \tag{10c}\\
		&\qquad\qquad\ \sum\limits_{i = 1}^t {\frac{{\gamma {\delta _t}{f_k}[i]{\kappa _k}}}{{{e^{\varepsilon /{\kappa _k}}} - {e^\varepsilon }}}}  \!\le\! \sum\limits_{i = 1}^t {{\bar R_k}[i]},\!\forall k \!\in\! {\cal K},\!\forall t \!\in\! {\cal T}\!,\!\label{eq10d} \tag{10d}\\
		&\qquad\qquad\ \text{(\ref{eq7g}), (\ref{eq7h})}.\label{eq10e} \tag{10e}
	\end{align}
\end{subequations}
(P3) is a convex problem that can be solved by convex programming tools, such as CVX \cite{grant2014cvx}. The uploading power $P_k[t]$ can be obtained by $P_k[t]=\frac{p_k[t]}{b_k[t]}$.

\subsection{Subproblem of Beamforming Designing}
We keep $\boldsymbol{X}$ and $\boldsymbol{Q}$ fixed in this subsection. The problem of designing $\boldsymbol{Y}$ can be formulated as
\begin{subequations}
	\label{eq11}
	\begin{align}
		&\text{(P4)}\mathop{\text{max}}\limits_{\boldsymbol{Y}}\ \sum\limits_{t = 1}^T {\sum\limits_{k = 1}^K {{\lambda_k}} {{R_k}[t]}} \label{eq11a} \tag{11a} \\
		&\qquad\ \text{s.t.}\quad \text{(\ref{eq7c}), (\ref{eq7d}), (\ref{eq7e}), (\ref{eq7f}),}\label{eq11b} \tag{11b}
	\end{align}
\end{subequations}
where the terms that are not related to $\boldsymbol{Y}$ in (\ref{eq11a}) are omitted. At each time instant, the BS and RIS only serve one UE due to the fact of that the downlink WPT and uplink uploading are scheduled in a TDM manner. A low-complexity alternating optimization (AO) algorithm is developed to design $\boldsymbol{Y}$.

\noindent\textbf{\emph{{Proposition }1: }}\emph{With the fixed RIS beamforming matrics, according to the maximum-ratio transmission (MRT) \cite{tse2005fundamentals}, the optimal transmit and receive beamforming vectors are}
\begin{equation}
	\label{eq12}
	\boldsymbol{w}_k^*[t]=\left\{  
	\begin{array}{lr}  
		\sqrt{P_{\text{T}}}\frac{\left( {{\boldsymbol{h}_k^H + \boldsymbol{g}_k^H[t]{{{\boldsymbol{\Theta}} }_k}[t]\boldsymbol{G}_{{\rm{UR}}}^H[t]}} \right)^H}{\left\Vert { {\boldsymbol{h}_k^H + \boldsymbol{g}_k^H[t]{{{\boldsymbol{\Theta}} }_k}[t]\boldsymbol{G}_{{\rm{UR}}}^H[t]}} \right\Vert},\!\! &a_k[t]>0,  \\  
		\boldsymbol{0}_{M\times1},\!\! & a_k[t]=0,  
	\end{array}  
	\right.
\end{equation}
and
\begin{equation}
	\label{eq13}
	\boldsymbol{u}_k^*[t]=\left\{  
	\begin{array}{lr}  
		\frac{\left( {{\boldsymbol{h}_k^H + \boldsymbol{g}_k^H[t]{{\bar{\boldsymbol{\Theta}} }_k}[t]\boldsymbol{G}_{{\rm{UR}}}^H[t]}} \right)^H}{\left\Vert { {\boldsymbol{h}_k^H + \boldsymbol{g}_k^H[t]{{\bar {\boldsymbol{\Theta}} }_k}[t]\boldsymbol{G}_{{\rm{UR}}}^H[t]}} \right\Vert}, &b_k[t]>0,  \\  
		\boldsymbol{0}_{M\times1}, & b_k[t]=0.
	\end{array}  
	\right.
\end{equation}
\emph{Proof:} Please refer to Section 3 of \cite{wu2019intelligent}.$\hfill\blacksquare$

\noindent\textbf{\emph{{Proposition }2: }}\emph{With the fixed BS beamforming vectors, the optimal value of the $n$-th RIS element serving the downlink WPT at the $t$-th time slot is}
\begin{equation}
	\label{eq14}
	\begin{aligned}
		&e^{j\theta_{k,n}^*[t]}=\\
		&\!\! \!\!\left\{ \!\! 
		\begin{array}{lr}  
			e^{\left\{j\theta_{k,0}[t]-j\arg\left(\boldsymbol{g}_{\text{UR},n}^H[t]\boldsymbol{w}_k[t]\right)-j\arg\left(g_{k,n}^H[t]\right)\right\}},\!\!\!\!\!\! &a_k[t]>0,  \\  
			0, \!\!\!\!\!\!& a_k[t]=0,  
		\end{array}  
		\right.
	\end{aligned}
\end{equation}
\emph{where $\theta_{k,0}[t]=\arg\left(\boldsymbol{h}_k^H\boldsymbol{w}_k[t]\right)$, $\boldsymbol{g}_{\text{UR},n}[t]$ denotes the $n$-th column of $\boldsymbol{G}_{{\rm{UR}}}[t]$, and $g_{k,n}[t]$ denotes the $n$-th element of $\boldsymbol{g}_k[t]$. The optimal value of the $n$-th RIS element serving the uploading at the $t$-th time slot is}
\begin{equation}
	\label{eq15}
	\begin{aligned}
		&e^{j\bar\theta_{k,n}^*[t]}=\\
		&\!\!\!\! \left\{ \!\! 
		\begin{array}{lr}  
			e^{\left\{j\bar\theta_{k,0}[t]-j\arg\left(\boldsymbol{g}_{\text{UR},n}^H[t]\boldsymbol{u}_k[t]\right)-j\arg\left(g_{k,n}^H[t]\right)\right\}},\!\!\!\!\!\! &b_k[t]>0,  \\  
			0, \!\!\!\!\!\!& b_k[t]=0,  
		\end{array}  
		\right.
	\end{aligned}
\end{equation}
\emph{where $\bar\theta_{k,0}[t]=\arg\left(\boldsymbol{h}_k^H\boldsymbol{u}_k[t]\right)$.}\\
\emph{Proof:} Please refer to Appendix A.$\hfill\blacksquare$

Using Proposition 1 and 2, we can obtain the optimal beamforming vectors and matrics by iteratively updating the BS beamforming vectors and the RIS beamforming matrics.

\subsection{Subproblem of UAV Trajectory Designing}
With the fixed $\boldsymbol{X}$ and $\boldsymbol{Y}$, the problem of designing the UAV trajectory can be written as
\begin{subequations}
	\label{eq16}
	\begin{align}
		&\text{(P5)}\mathop{\text{max}}\limits_{\boldsymbol{Q}}\ \sum\limits_{t = 1}^T {\sum\limits_{k = 1}^K {{\lambda_k}} {{R_k}[t]}} \label{eq16a} \tag{16a} \\
		&\qquad\ \text{s.t.}\quad \text{(\ref{eq7e}), (\ref{eq7f}), (\ref{eq7i}), (\ref{eq7j}),}\label{eq16b} \tag{16b}
	\end{align}
\end{subequations}
where the terms that are not related to $\boldsymbol{Q}$ in (\ref{eq16a}) are omitted. The phase shifts of BS-UEs links are aligned with those of BS-RIS-UEs links when the subproblem of designing $\boldsymbol{Y}$ is convergent. Hence, the amount of uploading data and the harvested energy of the $k$-th UE at the $t$-th time slot can be reformulated as
\begin{equation}
	\label{eq17}
	\begin{aligned}
		{\tilde R_k}[t] &= B{\delta _t}{b_k}[t]\\
		&\cdot{\log _2}\Bigg\{ 1 + \frac{P_k[t]}{{\sigma _\text{B}^2}}\bigg(\left|A_k[t]^2\right|+\frac{2\left|B_k[t]\right|h_0}{d_{\text{R2U},k}^{\alpha_{\text{RU}}/{2}}[t]d_{\text{B2R}}^{\alpha_{\text{BR}}/{2}}[t]}\\
		&\qquad\quad+\frac{\left|C_k[t]\right|^2h_0^2}{d_{\text{R2U},k}^{\alpha_{\text{RU}}}[t]d_{\text{B2R}}^{\alpha_{\text{BR}}}[t]}\bigg)\! \Bigg\}\!,\!\forall k\!\in\!\mathcal{K},
	\end{aligned}
\end{equation}
and
\begin{equation}
	\label{eq18}
	\begin{aligned}
	{\tilde E_k}[t] = {\eta _0}{\delta _t}\Bigg(&\left|D_k[t]\right|^2+\frac{2\left|F_k[t]\right|h_0}{d_{\text{R2U},k}^{\alpha_{\text{RU}}/{2}}[t]d_{\text{B2R}}^{\alpha_{\text{BR}}/{2}}[t]}\\
	&+\frac{\left|G_k[t]\right|^2h_0^2}{d_{\text{R2U},k}^{\alpha_{\text{RU}}}[t]d_{\text{B2R}}^{\alpha_{\text{BR}}}[t]}\Bigg){a_k}[t],\forall k\in\mathcal{K},
	\end{aligned}
\end{equation}
where $A_k[t]=\boldsymbol{h}_k^H\boldsymbol{u}_k[t]$, $D_k[t]=\boldsymbol{h}_k^H\boldsymbol{w}_k[t]$,
\begin{equation}
	\label{eq19}
	B_k[t]\!=\!\boldsymbol{h}_k^H\!{\boldsymbol{u}_k}[t]\boldsymbol{u}_k^H\![t]\boldsymbol{a}\left( {{\phi _0}\![t]} \right){\boldsymbol{b}^T}\!\!\left( {{\varphi _0}\![t]} \right)\bar {\boldsymbol{\Theta}} _k^H\![t]\boldsymbol{b}\left( {{\varphi _k}[t]} \right),
\end{equation}
\begin{equation}
	\label{eq20}
	C_k[t]={\boldsymbol{b}^H}\left( {{\varphi _k}[t]} \right){\bar {\boldsymbol{\Theta}} _k}[t]{\boldsymbol{b}^*}\left( {{\varphi _0}[t]} \right){\boldsymbol{a}^H}\left( {{\phi _0}[t]} \right){\boldsymbol{u}_k}[t],
\end{equation}
\begin{equation}
	\label{eq21}
	F_k[t]\!=\!\boldsymbol{h}_k^H\!{\boldsymbol{w}_k}[t]\boldsymbol{w}_k^H\![t]\boldsymbol{a}\left( {{\phi _0}\![t]} \right){\boldsymbol{b}^T}\!\!\left( {{\varphi _0}\![t]} \right)\!{\boldsymbol{\Theta}} _k^H\![t]\boldsymbol{b}\left( {{\varphi _k}[t]} \right),
\end{equation}
\begin{equation}
	\label{eq22}
	G_k[t]={\boldsymbol{b}^H}\left( {{\varphi _k}[t]} \right){ {\boldsymbol{\Theta}} _k}[t]{\boldsymbol{b}^*}\left( {{\varphi _0}[t]} \right){\boldsymbol{a}^H}\left( {{\phi _0}[t]} \right){\boldsymbol{w}_k}[t].
\end{equation}
Furthermore, (P5) can be reformulated as
\begin{subequations}
	\label{eq23}
	\begin{align}
		&\text{(P6)}\mathop{\text{max}}\limits_{\boldsymbol{Q},\atop \{x_k[t]\},\{y[t]\}}\ \sum\limits_{t = 1}^T \sum\limits_{k = 1}^K {{\lambda_k}}B{\delta _t}{b_k}[t]{{\log }_2}\Bigg\{1 + \frac{{{P_k}[t]}}{{\sigma _B^2}}\nonumber\\
		&\qquad\quad\  \left( {{{\left| {{A_k}[t]} \right|}^2} + \frac{{2\left| {{B_k}[t]} \right|{h_0}}}{{x_k^{1/2}[t]{y^{1/2}}[t]}} + \frac{{{{\left| {{C_k}[t]} \right|}^2}h_0^2}}{{{x_k}[t]y[t]}}} \right) \Bigg\} \label{eq23a} \tag{23a} \\
		&\ \text{s.t.}\ \sum\limits_{i = 1}^t {\left[ {\gamma {\delta _t}{\gamma _c}f_k^3[i] + {b_k}[i]{\delta _t}{P_k}[i]} \right]}\le\eta_0\delta_t\sum\limits_{i = 1}^t\Bigg(\left|D_k[i]\right|^2+ \nonumber\\
		&\quad\ \ \quad \frac{2\left|F_k[i]h_0\right|}{x_k^{1/2}[i]y^{1/2}[i]}+\frac{\left|G_k[i]\right|^2h_0^2}{x_k[t]y[t]}\Bigg),\forall k \!\in\! {\cal K},\forall t \!\in\! {\cal T},\label{eq23b} \tag{23b}\\
		&\quad\ \ \sum\limits_{i = 1}^t {\frac{{\gamma {\delta _t}{f_k}[i]{\kappa _k}}}{{{e^{\varepsilon /{\kappa _k}}} - {e^\varepsilon }}}}\le\sum\limits_{i = 1}^tB\delta_tb_k[i]{{\log }_2}\Bigg\{1 + \frac{{{P_k}[t]}}{{\sigma _B^2}}\bigg( {{\left| {{A_k}[i]} \right|}^2} \nonumber\\
		&\quad\ \ \quad\!+\!\frac{{2\left| {{B_k}[i]} \right|{h_0}}}{{x_k^{1/2}[i]{y^{1/2}}[i]}}\! + \!\frac{{{{\left| {{C_k}[i]} \right|}^2}h_0^2}}{{{x_k}[i]y[i]}} \bigg)\!\! \Bigg\},\forall k \!\in\!{\cal K},\forall t \!\in\! {\cal T}\!,\!\label{eq23c} \tag{23c}\\
		&\quad\ \ x_k^{1/\alpha_{\text{RU}}}\ge\left\Vert\boldsymbol{q}[t]-\boldsymbol{q}_k\right\Vert,\forall k \in {\cal K},\forall t \in {\cal T},\label{eq23d} \tag{23d}\\
		&\quad\ \ y^{\alpha_{\text{BR}}}\ge\left\Vert\boldsymbol{q}[t]-\boldsymbol{q}_0\right\Vert,\forall k \in {\cal K},\forall t \in {\cal T},\label{eq23e} \tag{23e}\\
		&\quad\ \ \text{(\ref{eq7i}), (\ref{eq7j})},\label{eq23f} \tag{23f}
	\end{align}
\end{subequations}
where $x_k[t],y[t],\forall k \in{\cal K},\forall t \in {\cal T}$ are slack variables. Due to the non-convex functions in (\ref{eq23a}), (\ref{eq23c}), and (\ref{eq23d}), (P6) is a non-convex problem. We propose a successive convex approximation (SCA) algorithm to solve (P6). The lower bounds of the uploading data rate and the harvested energy can be obtained using the first order Taylor expansion, which is given by (\ref{eq24}) and (\ref{eq25}) at the top of the next page. In (\ref{eq24}) and (\ref{eq25}), we have
\begin{figure*}[ht] 
	\begin{equation}
		\label{eq24}
		\begin{aligned}
			R_k[t]\le R_{\text{lb},k}[t]\triangleq& B\delta_tb_k[t]\Bigg\{\log_2\left(H_k^{(l)}[t]\right)\\
			&- \frac{{{P_k}[t]}}{{\sigma _B^2H_k^{(l)}[t]\ln 2}}\left( {\frac{{\left| {B_k^{(l)}[t]} \right|{h_0}}}{{{{\left( {x_k^{(l)}[t]} \right)}^{3/2}}{{\left( {{y^{(l)}}[t]} \right)}^{1/2}}}} + \frac{{{{\left| {C_k^{(l)}[t]} \right|}^2}h_0^2}}{{{{\left( {x_k^{(l)}[t]} \right)}^2}{y^{(l)}}[t]}}} \right) \cdot \left( {{x_k}[t] - x_k^{(l)}[t]} \right)\\
			&- \frac{{{P_k}[t]}}{{\sigma _B^2H_k^{(l)}[t]\ln 2}}\left( {\frac{{\left| {B_k^{(l)}[t]} \right|{h_0}}}{{{{\left( {x_k^{(l)}[t]} \right)}^{1/2}}{{\left( {{y^{(l)}}[t]} \right)}^{3/2}}}} + \frac{{{{\left| {C_k^{(l)}[t]} \right|}^2}h_0^2}}{{x_k^{(l)}[t]{{\left( {{y^{(l)}}[t]} \right)}^2}}}} \right)\cdot{\left( {y[t] - {y^{(l)}}[t]} \right)}\Bigg\},
		\end{aligned}
	\end{equation}
\vspace{-1.5em}
\end{figure*}
\begin{figure*}[ht] 
	\begin{equation}
		\label{eq25}
		\begin{aligned}
			{E_k}[i] \ge {E_{{\rm{lb}},k}}[i]\triangleq&{\eta _0}{\delta _t}{a_k}[i]\left\{ {U_k^{(l)}[i] - \left( {\frac{{\left| {F_k^{(l)}[i]} \right|{h_0}}}{{{{\left( {x_k^{(l)}[i]} \right)}^{3/2}}{{\left( {{y^{(l)}}[i]} \right)}^{1/2}}}} + \frac{{{{\left| {G_k^{(l)}[i]} \right|}^2}h_0^2}}{{{{\left( {x_k^{(l)}[i]} \right)}^2}{y^{(l)}}[i]}}} \right) \cdot \left( {{x_k}[i] - x_k^{(l)}[i]} \right)} \right.\\
			&- \left. {\left( {\frac{{\left| {F_k^{(l)}[i]} \right|{h_0}}}{{{{\left( {x_k^{(l)}[i]} \right)}^{1/2}}{{\left( {{y^{(l)}}[i]} \right)}^{3/2}}}} + \frac{{{{\left| {G_k^{(l)}[i]} \right|}^2}h_0^2}}{{x_k^{(l)}[i]{{\left( {{y^{(l)}}[i]} \right)}^2}}}} \right) \cdot \left( {y[i] - {y^{(l)}}[i]} \right)} \right\},
		\end{aligned}
	\end{equation}
	\hrulefill  
\end{figure*}
\begin{equation}
	\label{eq26}
	\begin{aligned}
	&H_k^{(l)}[t]= \\
	&\left( {1 + \frac{{{P_k}[t]}}{{\sigma _B^2}}\left( {{{\left| {{A_k}[t]} \right|}^2} + \frac{{2\left| {{B_k}[t]} \right|{h_0}}}{{\sqrt {x_k^{(l)}[t]{y^{(l)}}[t]} }} + \frac{{{{\left| {{C_k}[t]} \right|}^2}h_0^2}}{{x_k^{(l)}[t]{y^{(l)}}[t]}}} \right)} \right),
	\end{aligned}
\end{equation}
and
\begin{equation}
	\label{eq27}
	U_k^{(l)}[i] = {\left| {{D_k}[i]} \right|^2} + \frac{{2\left| {F_k^{(l)}[i]} \right|{h_0}}}{{\sqrt {x_k^{(l)}[i]{y^{(l)}}[i]} }} + \frac{{{{\left| {G_k^{(l)}[i]} \right|}^2}h_0^2}}{{x_k^{(l)}[i]{y^{(l)}}[i]}},
\end{equation}
where $l$ denotes the number of SCA iteration. Finally, the problem of designing the UAV trajectory is formulated as
\begin{subequations}
	\label{eq28}
	\begin{align}
		&\text{(P7)}\mathop{\text{max}}\limits_{\boldsymbol{Q},\atop \{x_k[t]\},\{y[t]\}}\ \sum\limits_{t = 1}^T {\sum\limits_{k = 1}^K {{\lambda_k}} {R_{\text{lb},k}}[t]} \label{eq28a} \tag{28a} \\
		&\qquad\quad\ \ \text{s.t.}\quad\sum\limits_{i = 1}^t {\left[ {\gamma {\delta _t}{\gamma _c}f_k^3[i] + {b_k}[i]{\delta _t}{P_k}[i]} \right]} \nonumber\\
		&\qquad\qquad\qquad\quad\le \sum\limits_{i = 1}^t{E_{\text{lb},k}[i]},\forall k \in {\cal K},\forall t \in {\cal T},\label{eq28b} \tag{28b}\\
		&\quad\qquad\qquad\ \ \sum\limits_{i = 1}^t {\frac{{\gamma {\delta _t}{f_k}[i]{\kappa _k}}}{{{e^{\varepsilon /{\kappa _k}}} - {e^\varepsilon }}}}  \nonumber\\
		&\qquad\qquad\qquad\quad\le \sum\limits_{i = 1}^t {{R_{\text{lb},k}}[i]},\forall k \in {\cal K},\forall t \in {\cal T},\label{eq28c} \tag{28c}\\
		&\quad\qquad\qquad\ \ \text{(\ref{eq23d}), (\ref{eq23e}), (\ref{eq23f})}.\label{eq28d} \tag{28d}
	\end{align}
\end{subequations}
(P7) is a standard convex problem that can be solved by CVX tools. The SCA algorithm for designing the UAV trajectory is presented in \textbf{Algorithm 1}. The overall BCD-based algorithm for designing $\boldsymbol{X}$, $\boldsymbol{Y}$, and $\boldsymbol{Q}$ is presented in \textbf{Algorithm 2}. In Algorithms 1 and 2, $O^{(i)}$ and $\bar O^{(j)}$ denote the value of objective function in (\ref{eq7a}).
\begin{algorithm}[t]
	\caption{SCA algorithm for designing the UAV trajectory.}\label{alg:alg1}
	\small
	\begin{algorithmic}
		\STATE 
		\STATE 1. $ \textbf{Inputs: }$$\boldsymbol{X}$, $\boldsymbol{Y}$, $K$, $T$, $\lambda_k$, $\boldsymbol{q}_k$, $\boldsymbol{q}_0$, $\boldsymbol{q}_1$, $\boldsymbol{q}_T$, $\delta_t$, $z$, $z_{\text{B}}$, $z_k$, $V_{\text{max}}$, $\eta_0$, $B$, $\sigma_{\text{B}}^2$, $h_0$, $K\alpha_{\text{BR}}$, $\alpha_{\text{BU}}$, $\alpha_{\text{RU}}$, $\gamma$, $\gamma_c$, $\varepsilon$, $\epsilon$, $i_{\text{max}}$. 
		\STATE 2. $ \textbf{Outputs: }$UAV trajectory $\boldsymbol{Q}$ for (P5).
		\STATE 3. $ \textbf{Initialization: }$Solve (P7) without considering the constraints in (\ref{eq28b}) and (\ref{eq28c}) to obtain $\boldsymbol{Q}^{(0)}$, which represent the initial UAV trajectory, and calculate the amount of total uploading data as $O_{0}$. Set $i=1$ and $O_{1}=+\inf$. 
		\WHILE {$i\leq i_{\text{max}}$ and $\frac{\vert O_{i}-O_{i-1}\vert}{\vert O_{i-1}\vert}\geq\epsilon$}
		\STATE 4. Calculate the constant terms of (\ref{eq24}) and (\ref{eq25}) using $\boldsymbol{Q}^{(i-1)}$.
		\STATE 5. Solve (P7) to obtain $\boldsymbol{Q}^{(i)}$ and $O_{i}$.
		\ENDWHILE
		\STATE 6. Set $\boldsymbol{Q}=\boldsymbol{Q}^{(i)}$.
	\end{algorithmic}
	\label{alg1}
\end{algorithm}
\begin{algorithm}[t]
	\caption{BCD algorithm for solving (P1).}\label{alg:alg2}
	\small
	\begin{algorithmic}
		\STATE 
		\STATE 1. $ \textbf{Inputs: }$$P_{\text{T}}$, $\kappa_k$, $K$, $T$, $\lambda_k$, $\boldsymbol{q}_k$, $\boldsymbol{q}_0$, $\boldsymbol{q}_1$, $\boldsymbol{q}_T$, $\delta_t$, $z$, $z_{\text{B}}$, $z_k$, $V_{\text{max}}$, $\eta_0$, $B$, $\sigma_{\text{B}}^2$, $h_0$, $K\alpha_{\text{BR}}$, $\alpha_{\text{BU}}$, $\alpha_{\text{RU}}$, $\gamma$, $\gamma_c$, $\varepsilon$, $\epsilon$, $i_{\text{max}}$, $j_{\text{max}}$. 
		\STATE 2. $ \textbf{Outputs: }$$\boldsymbol{X}$, $\boldsymbol{Y}$, $\boldsymbol{Q}$ for (P1).
		\STATE 3. $ \textbf{Initialization: }$Random initialize $\boldsymbol{w}_k^{(1)}$ and $\boldsymbol{u}_k^{(1)}$. Set $\boldsymbol{\Theta}_k^{(1)}$ and $\bar{\boldsymbol{\Theta}}_k^{(1)}$ according to Proposition 2. Set $\boldsymbol{Q}^{(1)}$ according to that the UAV flies straight to $\left[x_{\text{B}},y_{\text{B}},z\right]^T$ and then flies straight to $\boldsymbol{q}_T$. Set $i=1$, $O_{0}=-\inf$, and $O_{1}=+\inf$. 
		\WHILE {$i\leq i_{\text{max}}$ and $\frac{\vert O_{i}-O_{i-1}\vert}{\vert O_{i-1}\vert}\geq\epsilon$}
		\STATE 4. Solve (P3) to obtain $\boldsymbol{X}^{(i)}$.
		\STATE 5. Set $j=1$, $\bar O_{0}=-\inf$, and $\bar O_{1}=+\inf$.
		\WHILE {$j\leq j_{\text{max}}$ and $\frac{\vert \bar O_{j}-\bar O_{j-1}\vert}{\vert \bar O_{j-1}\vert}\geq\epsilon$}
		\STATE 6. Obtain $\boldsymbol{w}_k^{(j)}$ and $\boldsymbol{u}_k^{(j)}$ by Proposition 1.
		\STATE 7. Obtain $\boldsymbol{\Theta}_k^{(j)}$ and $\bar{\boldsymbol{\Theta}}_k^{(j)}$ by Proposition 3. Calculate $\bar O_{j}$ using the updated beamforming vectors and matrics.
		\ENDWHILE
		\STATE 8. Set ${\boldsymbol{\Theta}}_k^{(i)}={\boldsymbol{\Theta}}_k^{(j)}$ and $\bar{\boldsymbol{\Theta}}_k^{(i)}=\bar{\boldsymbol{\Theta}}_k^{(j)}$.
		\STATE 9. Solve (P5) using Algorithm 1 to obtain $\boldsymbol{Q}^{(i)}$ and $O_{i+1}$.
		\ENDWHILE
		\STATE 10. Set $\boldsymbol{X}=\boldsymbol{X}^{(i)}$, $\boldsymbol{Y}=\boldsymbol{Y}^{(i)}$, and $\boldsymbol{Q}=\boldsymbol{Q}^{(i)}$.
	\end{algorithmic}
	\label{alg2}
\end{algorithm}

\section{Uploading Data Maximization Based on Lossy Compression}
The problem of the uploading data maximization based on lossy compression can be formulated as
\begin{subequations}
	\label{eq29}
	\begin{align}
		&\text{(P8)}\mathop{\text{max}}\limits_{\bar{\boldsymbol{X}},\boldsymbol{Y},\boldsymbol{Q}}\ \sum\limits_{t = 1}^T {\sum\limits_{k = 1}^K {{\lambda_k}} \left\{ {{R_k}[t] + \left( {1 - {\kappa _k}} \right){\bar S_k}[t]} \right\}} \label{eq29a} \tag{29a} \\
		&\qquad\ \ \text{s.t.}\quad{\bar f_k}[t],{P_k}[t],{a_k}[t],{b_k}[t] \ge 0,\forall k\! \in\! {\cal K},\forall t \!\in\! {\cal T}\!,\!\label{eq29b} \tag{29b} \\
		&\qquad\qquad\ \ \sum\limits_{i = 1}^t {\left[ {\gamma {\delta _t}{\gamma _c}\bar f_k^3[i] + {b_k}[i]{\delta _t}{P_k}[i]} \right]} \nonumber\\
		&\qquad\qquad\qquad\le \sum\limits_{i = 1}^t{E_k[i]},\forall k \in {\cal K},\forall t \in {\cal T},\label{eq29c} \tag{29c}\\
		&\qquad\qquad\ \ \sum\limits_{i = 1}^t {\frac{{\gamma {\delta _t}{\bar f_k}[i]{\kappa _k}}}{{{e^{\varepsilon /{\kappa _k}}} - {e^\varepsilon }}}}  \!\le\! \sum\limits_{i = 1}^t {{R_k}[i]},\!\forall k \!\in\! {\cal K},\!\forall t \!\in\! {\cal T}\!,\!\label{eq29d} \tag{29d}\\
		&\qquad\qquad\ \ \text{(\ref{eq7c}), (\ref{eq7d}), (\ref{eq7g}), (\ref{eq7h}), (\ref{eq7i}), (\ref{eq7j})},\label{eq29e} \tag{29e}
	\end{align}
\end{subequations}
where $\bar{\boldsymbol{X}}\triangleq\left\{ {{\bar f_k}[t],{P_k}[t],{a_k}[t],{b_k}[t],\forall k \in {\cal K},\forall t \in {\cal T}} \right\}$. Similar with (P1), (P8) can also be solved by the BCD algorithm. Therfore, we omit the details of solving (P8).

\section{Analysis of Energy Efficiency}
The essence of the fact that compression can improve system performance lies in its enhancement of energy efficiency. Therefore, the energy efficiencies of the compression and uploading are analyzed in this section. The compression schemes can improve system performance when the power used for compression is relatively small. However, the benefits of the compression schemes gradually reduce when the power used for compression becomes relatively large.

The energy efficiency of directly uploading data is defined as
\begin{equation}
	\label{eq30}
	\begin{aligned}
		&\eta_{\text{U}}\triangleq\frac{R_k[t]}{P_k[t]}=B{\delta _t}{b_k}[t]\\
		&\cdot\!{\log _2}\!\left( {\!1\!\! + \!\!\frac{{{{\left| {\left( {\boldsymbol{h}_k^H + \boldsymbol{g}_k^H[t]{{\bar {\boldsymbol{\Theta}} }_k}[t]\boldsymbol{G}_{{\rm{UR}}}^H[t]} \right){\boldsymbol{u}_k}[t]} \right|}^2}{P_k}[t]}}{{\sigma _\text{B}^2}}}\! \right)/P_k[t].
	\end{aligned}
\end{equation}
The energy efficiencies of the lossless and lossy compressions are defined as
\begin{equation}
	\label{eq31}
	\begin{aligned}
	{\eta _{\rm{C}}} \triangleq \frac{{\left( {1 - {\kappa _k}[t]} \right){S_k}[t]}}{{{E_{{\rm{C}},k}}[t]}} &= \frac{{\left( {1 - {\kappa _k}} \right){f_k}[t]}}{{\left( {{e^{\varepsilon /{\kappa _k}}} - {e^\varepsilon }} \right){P_k}[t]}} \\
	&= \frac{{\left( {1 - {\kappa _k}} \right)}}{{\gamma _c^{1/3}\left( {{e^{\varepsilon /{\kappa _k}}} - {e^\varepsilon }} \right)P_k^{2/3}[t]}},
	\end{aligned}
\end{equation}
and
\begin{equation}
	\label{eq32}
	\begin{aligned}
		{\bar \eta _{\rm{C}}} = \frac{{\left( {1 - {{\kappa }_k}} \right){{\bar S}_k}[t]}}{{\gamma {\delta _t}{\gamma _c}\bar f_k^3[t]}} &= \frac{{\sqrt {{{\kappa }_k}} \left( {1 - {{\kappa }_k}} \right){{\bar f}_k}[t]}}{{\left( {{e^{\varepsilon /{{\kappa }_k}}} - {e^\varepsilon }} \right){P_k}[t]}} \\
		&= \frac{{\sqrt {{{\kappa }_k}} \left( {1 - {{\kappa }_k}} \right)}}{{\gamma _c^{1/3}\left( {{e^{\varepsilon /{{\kappa }_k}}} - {e^\varepsilon }} \right)P_k^{2/3}[t]}}.
	\end{aligned}
\end{equation}
\noindent\textbf{\emph{{Proposition }3: }}\emph{There exists $P^*$ that makes $\eta_{\text{U}}<\eta_{\rm{C}}$ when $P_k[t]<P^*$, $\eta_{\text{U}}\ge\eta_{\rm{C}}$ when $P_k[t]\ge P^*$}

\emph{Proof:} Please refer to Appendix B.$\hfill\blacksquare$

\noindent\textbf{\emph{{Proposition }4: }}\emph{There exists $\bar P^*$ that makes $\eta_{\text{U}}<\bar\eta_{\rm{C}}$ when $P_k[t]<\bar P^*$, $\eta_{\text{U}}\ge\bar\eta_{\rm{C}}$ when $P_k[t]\ge \bar P^*$}

\emph{Proof:} The proof is omitted since it is similar to that of Proposition 3.$\hfill\blacksquare$
\begin{figure}[!t]
	\centering
	\includegraphics[width=0.5\textwidth]{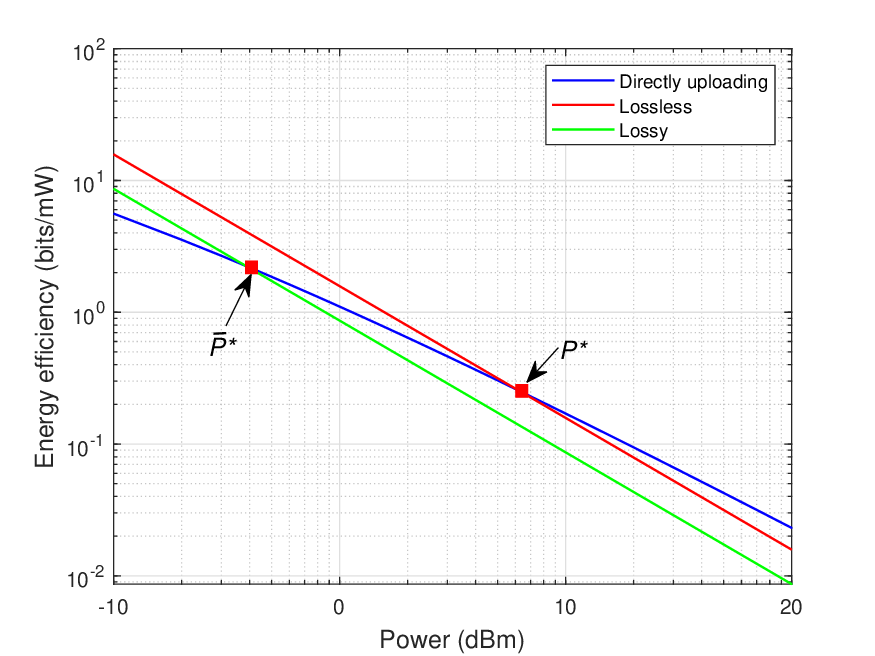}
	\caption{Energy efficiency versus the power.}
	\label{fig2}
\end{figure}

To illustrate the insight of Proposition 3 and 4, we observe the energy efficiency versus the power used for compression and uploading in Fig. \ref{fig2}, where the energy efficiency of the direct uploading scheme gradually becomes bigger than the lossless and lossy compression schemes. Moreover, $\bar P^* < P^*$ is because the energy efficiency of the lossless compression scheme is bigger than that of the lossy scheme.

\section{Analysis of Complexity and Convergence Performance}
\subsection{Analysis of Computational Complexity}
We mainly focus on the complexity analysis of solving the convex optimization problem, i.e., the complexity of solving (P3) and (P7). (P3) and (P7) can both be solved by the interior-point method, whose complexity is $\mathcal{O}(n^{3.5})$ with $n$ denoting the dimension of the optimization variables \cite{qi2024uav}. Therefore, the computational complexity of solving (P3) and (P7) are $\mathcal{O}((4KT)^{3.5})$ and $\mathcal{O}((2T)^{3.5})$, respectively. The complexity of Algorithm 1 is $\mathcal{O}(m(2T)^{3.5})$ with $m$ denoting the number of SCA iteration. The complexity of Algorithm 2 is $\mathcal{O}(n(\left(4KT\right)^{3.5}+KT\left(24MN+16N^2+4M\right)+m\left(2T\right)^{3.5}))$ with $n$ denoting the number of BCD iteration and $KT\left(24MN+16N^2+4M\right)$ denoting the complexity to obtain the optimal beamforming vectors and matrics. 

\subsection{Analysis of Convergence Performance}
Denoting the sum of uploading data at the $i$-th BCD iteration by $R\left(\boldsymbol{X}[i],\boldsymbol{Y}[i],\boldsymbol{Q}[i]\right)$. We have
\begin{equation}
	\label{eq33}
	\begin{aligned}
		R_{\text{max}}&\overset{\text{(a)}}{\geq}R\left(\boldsymbol{X}[i+1],\boldsymbol{Y}[i+1],\boldsymbol{Q}[i+1]\right)\\
		&\overset{\text{(b)}}{\geq}R\left(\boldsymbol{X}[i+1],\boldsymbol{Y}[i+1],\boldsymbol{Q}[i]\right)\\
		&\overset{\text{(c)}}{\geq}\!R\left(\boldsymbol{X}[i+1],\boldsymbol{Y}[i],\boldsymbol{Q}[i]\right)\!\overset{\text{(d)}}{\geq}\!R\left(\boldsymbol{X}[i],\boldsymbol{Y}[i],\boldsymbol{Q}[i]\right)\!,
	\end{aligned}
\end{equation}
where $R_{\text{max}}$ denotes the upper bound of the sum of uploading data, (b) holds since the Taylor expansion in (\ref{eq24}) is tight and (P7) can be optimally solved, (c) holds since Proposition 1 and 2 guarantee the uploading data non-decreasing, and (d) holds since the optimal $\boldsymbol{X}[i+1]$ can be obtained by solving (P3). Therefore, the convergence of the proposed BCD algorithm is guaranteed.

\section{Simulation Results}
\begin{table}[!t]
	\centering
	\caption{Parameter values in our simulations.}
	\label{table:1}
	\begin{tabular}{|l| l| l|}
		\hline
		$M$ & Number of BS antennas & 16 \\ 
		\hline
		$P_{\text{T}}$ & Transmit power of BS & 35dBm \\
		\hline
		$K$ & Number of UEs & 4 \\
		\hline
		$\{\lambda_k\}$ & UE weights & $\{0.1,0.1,0.4,0.4\}$ \\
		\hline
		$z_k,z_{\text{B}}$ & Heights of UEs and BS & 0 \\
		\hline
		$z$ & UAV height & 8m \\
		\hline
		$\{\boldsymbol{q}_0\}$ & Coordinate of BS & $(0,0,0)$ \\
		\hline
		$\{\boldsymbol{q}_k\}$ & Coordinates of UE & \makecell[l]{\{(-10m, 10m, 0), (-10m, 0, 0),\\ (10m, 0, 0), (10m, 10m, 0)\}} \\
		\hline
		$\boldsymbol{q}_1$ & UAV initial position & (-10m, 10m, 8m)\\
		\hline
		$\boldsymbol{q}_T$ & UAV final position & (10m, 10m, 8m)\\
		\hline
		$T$ & Number of time slots & 50 \\
		\hline
		$\delta_t$ & Time duration of each slot & 0.04s \\
		\hline
		$V_{\text{max}}$ & Maximum speed of UAV & 8m/s \\
		\hline
		$\eta_0$ & \makecell[l]{Energy conservation \\efficiency} & 0.8 \\
		\hline
		$B$ & System bandwidth & 40MHz \\
		\hline
		$\sigma_{\text{B}}^2$ & Noise variance & -60dBm \\
		\hline
		$h_0$ & Channel gain of 1m & -50dB \\
		\hline
		$\alpha_{\text{BR}}$ & Pathloss exponents & 2 \\
		\hline
		$\alpha_{\text{BU}}$ & Pathloss exponents & 4 \\
		\hline
		$\alpha_{\text{RU}}$ & Pathloss exponents & 2 \\
		\hline
		$\gamma$ & Compression time ratio & 0.8 \\
		\hline
		$\gamma_c$ & Hardware constant & $10^{-7}$ \\
		\hline
		$\varepsilon$ & Compression constant & $1.38$ \\
		\hline
	\end{tabular}
\end{table}
We present simulation results in this section to demonstrate the effectiveness of the proposed schemes. If not otherwise specified, the parameter values used in our simulations are listed in Table \uppercase\expandafter{\romannumeral1}.
\begin{figure}[!t]
	\centering
	\includegraphics[width=0.5\textwidth]{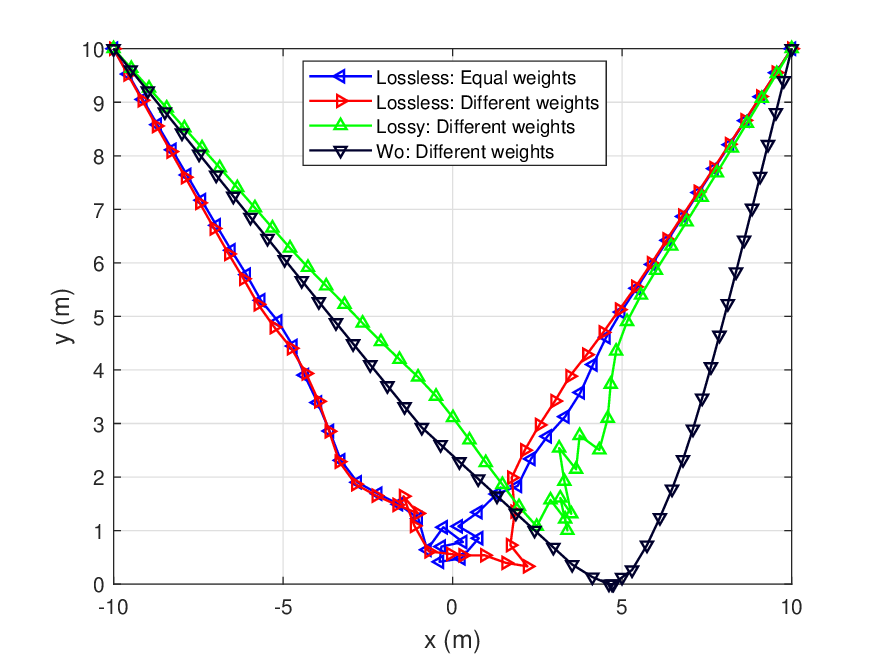}
	\caption{The UAV trajectories under different schemes.}
	\label{fig3}
\end{figure}

Fig. \ref{fig3} shows the UAV trajectories obtained by the lossless compression scheme, the lossy compression scheme, and the scheme without compression, which are respectively labelled as ``Lossless'', ``Lossy'', and ``Wo''. First, under the lossless scheme with equal UE weights, the UAV flies towards the BS and then hovers around the BS. The trajectory is nearly symmetrical about the straight line $x=0$. This is intuitive because a symmetric trajectory provides equal WPT and uploading opportunities for UEs with equal weights. Then, the trajectory under the lossless scheme with different weights tends to be close to the UEs with higher weights since the UAV can transfer more energy to these UEs and collect more data from them under such a trajectory. Moreover, the trajectories under the lossy scheme and the without compression scheme with different weights are closer to the UEs with higher weights than that of the lossless scheme. This is because the UAV needs to fly close to these UEs to provide the same level of transmission efficiency as the lossless scheme.

\begin{figure}[!t]
	\centering
	\includegraphics[width=0.5\textwidth]{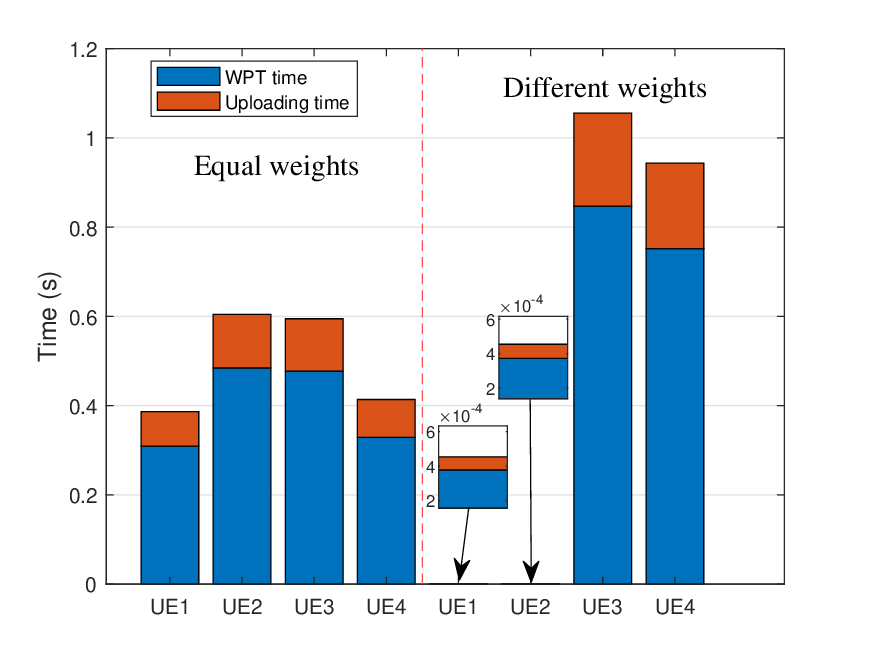}
	\caption{The slots allocation for WPT and uploading during the UAV flight under the lossless compression scheme.}
	\label{fig4}
\end{figure}

Fig. \ref{fig4} shows the time allocation of UEs with equal and different weights. The WPT time of each UE is longer than their uploading time. This is because WPT can benefit simultaneously from both BS and RIS beamforming gains, whereas uploading can only leverage the RIS beamforming gain. Therefore, more energy needs to be transmitted to the UEs for them to upload data. Moreover, the WPT and uploading times for UE2 and UE3 are longer than those for UE1 and UE4. This is because the distances between UE2 and the BS and between UE3 and the BS are shorter than those distances of UE1 and UE4, and the path losses of the channels associated with UE2 and UE3 are smaller than those of UE1 and UE4. Additionally, when the UE weights are different, the time slots are almost entirely allocated to the UEs with higher weights to maximize the amount of data uploaded.

\begin{figure}[!t]
	\centering
	\includegraphics[width=0.5\textwidth]{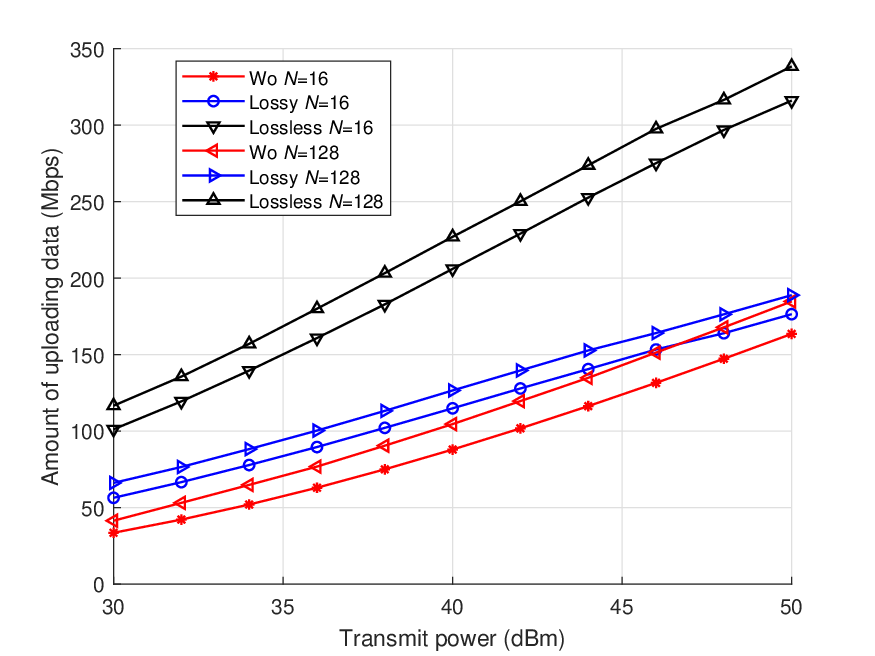}
	\caption{The amount of uploading data versus the transmit power of BS.}
	\label{fig5}
\end{figure}

Fig. \ref{fig5} shows the amount of uploading data of all schemes versus the transmit power of the BS when $N=16$ and $N=128$. The amount of uploading data of all schemes increases with the transmit power, as the UEs can harvest more energy and upload more data with higher transmit power. The slope of the lossless scheme is steeper than the other two schemes because of its higher energy efficiency. Moreover, the amount of uploading data of $N=128$ is greater than that of $N=16$ because more RIS elements can provide higher beamforming gain. Furthermore, with a fixed number of RIS elements, the uploading data of the lossless scheme is greater than that of the lossy scheme, which is greater than that of the without compression scheme. Additionally, the amount of uploading data of the without compression scheme gradually approaches that of the lossy scheme as the transmit power increases. This is because, with the increase in transmit power, the energy efficiency of the lossy scheme gradually becomes lower than that of the without compression scheme, leading to a convergence in the amount of uploaded data. This is consistent with Proposition 3.

\begin{figure}[!t]
	\centering
	\includegraphics[width=0.5\textwidth]{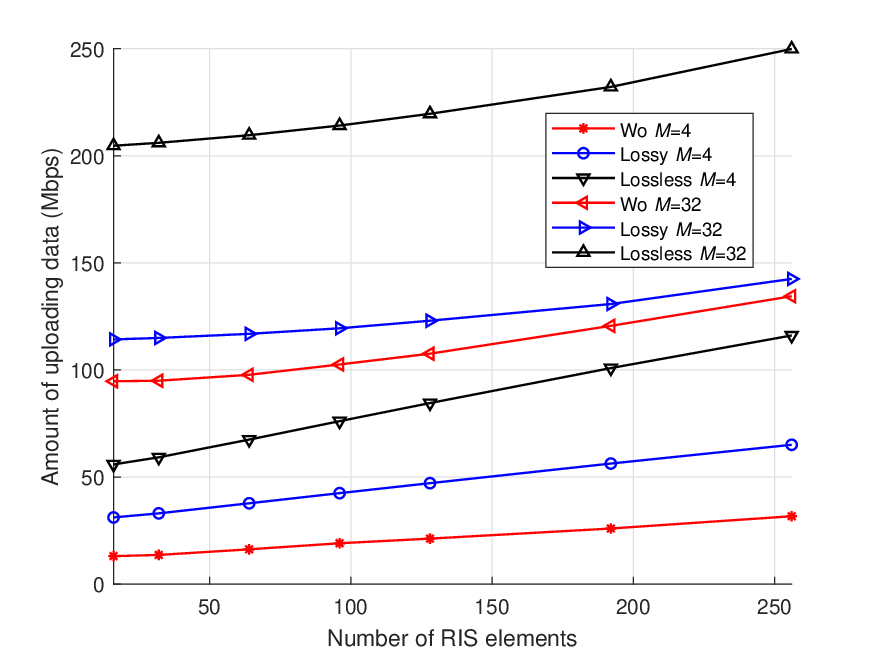}
	\caption{The amount of uploading data versus the number of RIS elements.}
	\label{fig6}
\end{figure}

Fig. \ref{fig6} shows the amount of uploading data of all schemes versus the number of RIS elements when $M=4$ and $M=32$. The amount of uploading data of all schemes increase with the number of RIS elements, as the increasing RIS elements can provide higher beamforming gain. Moreover, the amount of uploading data of $M=32$ is greater than that of $M=4$ because more BS antennas can provide higher degrees of freedom (DoF). Additionally, with a fixed number of BS antennas, the uploading data of the lossless scheme is greater than that of the lossy scheme, which is greater than that of the without compression scheme.

\begin{figure}[!t]
	\centering
	\includegraphics[width=0.5\textwidth]{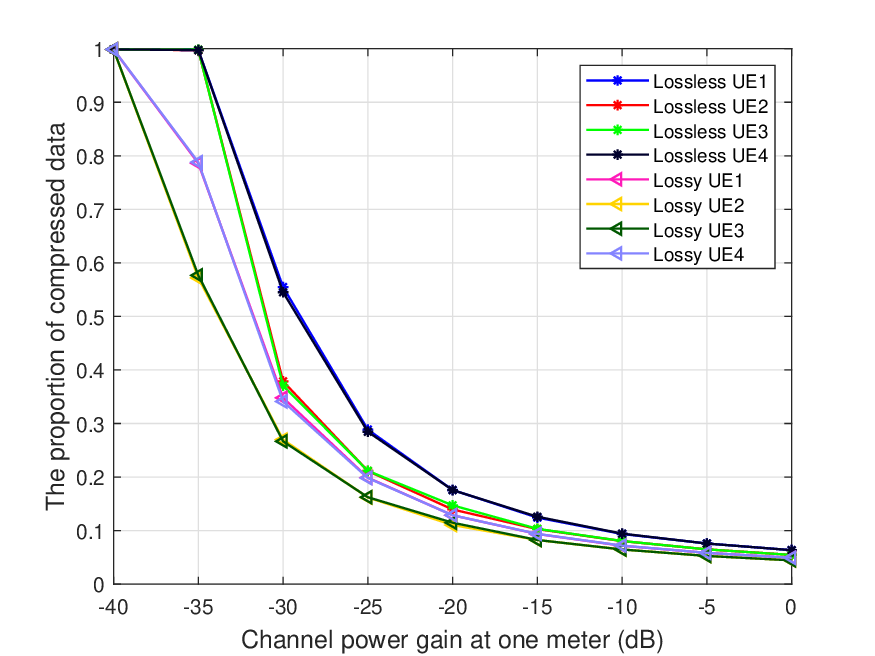}
	\caption{The proportion of compressed data in uploading data.}
	\label{fig7}
\end{figure}

Fig. \ref{fig7} shows the proportion of compressed data in uploading data versus the channel power gain when the UAV hovers at (0m, 0m, 8m). The proportions of compressed data of both the lossless scheme and lossy scheme decrease as the channel power gain increases. This is because higher proportions of compressed data can improve the transmission efficiency and overcome the effects of poor channel conditions. Additionally, the lossy scheme has lower proportions of compressed data than the lossless scheme since some compressed data are lost due to the compression error. Furthermore, in each compression scheme, the proportions of UE1 and UE4 are the same, and the proportions of UE2 and UE3 are also the same, since the path losses of UE1 and UE4 are the same, and the path losses of UE2 and UE3 are also the same. The proportions of compressed data of UE1 and UE4 are higher than those of UE2 and UE3 since the path losses of UE1 and UE4 are more serious. UE1 and UE4 need to enhance their transmission efficiencies by compressing more data.

\begin{figure}[!t]
	\centering
	\includegraphics[width=0.5\textwidth]{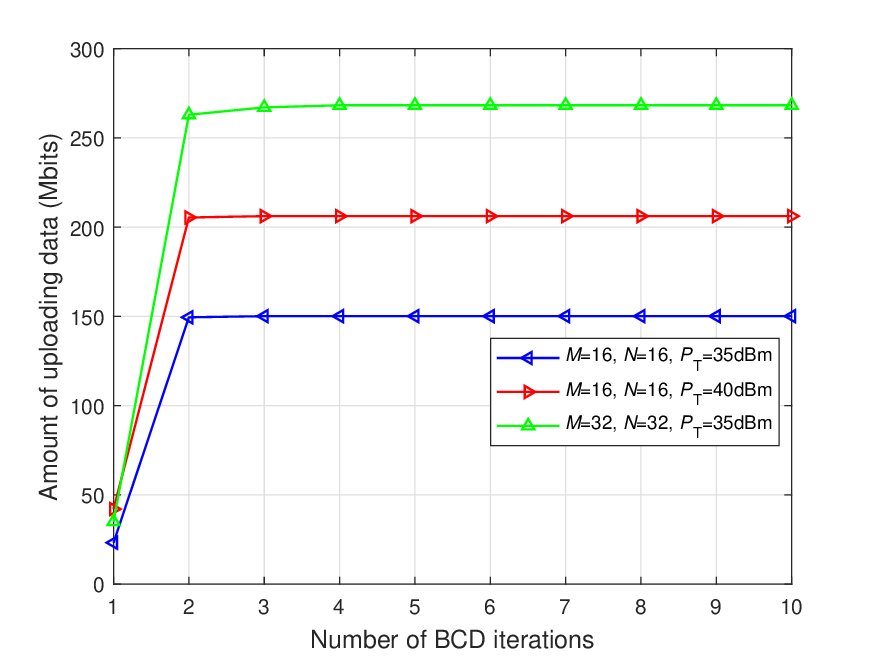}
	\caption{The convergence performance of the proposed BCD algorithm.}
	\label{fig8}
\end{figure}

Fig. \ref{fig8} demonstrates that the proposed BCD algorithm can achieve convergence with ten BCD iterations under different parameter configurations. The fast convergence is because we can obtain the globally optimal $\boldsymbol{X}$ by solving (P3) and the nearly globally optimal $\boldsymbol{Y}$ by Proposition 1 and 2, and the Taylor expansions used for dealing with (P6) construct tight lower bounds for the uploading data and the harvest energy.

\section{Conclusion}
In this paper, we have proposed a novel transmission framework for the WPT-assisted MCS systems enhanced by UAV-mounted RISs to enhance transmission efficiency. Two optimization problems are established to maximize the weighted sum of uploading data under different compression schemes. A BCD algorithm has been proposed to decompose the optimization problem into three subproblems, which are solved alternately. The subproblem for designing the transmission parameters can be easily converted into a convex problem. The closed-form optimal beamforming vectors and matrices are obtained when solving the beamforming subproblem. We utilize the SCA algorithm to approximate the lower bounds of the harvested energy by UEs and the uploading data for designing the UAV trajectory. The energy efficiencies of the compression schemes have been analyzed to further evaluate the gains obtained by compression. We have also analyzed the complexity and the convergence performance of the proposed algorithms. Numerous simulation results demonstrate that the proposed transmission framework and algorithms can effectively increase the amount of the collected data.

\section*{Appendix A\\Proof of Proposition 2}
In the expression of the harvested energy (i.e., (\ref{eq1})), the term that is related to the RIS beamforming matrix can be expressed as
\begin{equation}
	\label{eqa1}
	\begin{aligned}
	&\left| {\left( {\boldsymbol{h}_k^H + \boldsymbol{g}_k^H[t]{\boldsymbol{\Theta}_k}[t]\boldsymbol{G}_{{\rm{UR}}}^H[t]} \right){\boldsymbol{w}_k}[t]} \right|\\
	&=\left| { {\boldsymbol{h}_k^H{\boldsymbol{w}_k}[t] + \boldsymbol{g}_k^H[t]{\boldsymbol{\Theta}_k}[t]\boldsymbol{G}_{{\rm{UR}}}^H[t]} {\boldsymbol{w}_k}[t]} \right|\\
	&\overset{\text{(1)}}{\leq}\left| \boldsymbol{h}_k^H{\boldsymbol{w}_k}[t] \right| +\left| \boldsymbol{g}_k^H[t]{\boldsymbol{\Theta}_k}[t]\boldsymbol{G}_{{\rm{UR}}}^H[t] {\boldsymbol{w}_k}[t] \right|,
	\end{aligned}
\end{equation}
where (1) holds due to the triangle inequality. When $\arg\left(\boldsymbol{h}_k^H{\boldsymbol{w}_k}[t]\right)=\arg\left(\boldsymbol{g}_k^H[t]{\boldsymbol{\Theta}_k}[t]\boldsymbol{G}_{{\rm{UR}}}^H[t] {\boldsymbol{w}_k}[t]\right)$, the equality holds. The RIS beamforming matrix is used to align the phase shift of the reflected link with the phase shift of the direct link. The phase shift difference between the two links is $\arg\left(\boldsymbol{h}_k^H{\boldsymbol{w}_k}[t]\right)-\arg\left(\boldsymbol{g}_{\text{UR},n}^H[t]\boldsymbol{w}_k[t]\right)-\arg\left(g_{k,n}^H[t]\right)$. Then, we can obtain the optimal values of the RIS elements in (\ref{eq14}). We omit the derivation of (\ref{eq15}) since it is similar to that of (\ref{eq15}). This completes the proof of Proposition 2.

\section*{Appendix B\\Proof of Proposition 3}
We first check the first-order partial derivatives of $\eta_{\text{U}}$ and $\eta_{\text{C}}$, which are expressed in (\ref{eqa2}) and (\ref{eqa3}). (\ref{eqa2}) is shown at the top of this page.
\begin{figure*}[ht] 
	\begin{equation}
	\label{eqa2}
	\frac{\partial\eta_{\text{U}}}{\partial P_k[t]}=
	\frac{B\delta_tb_k[t]\left[U_k[t]P_k[t]/\sigma_{\text{B}}^2-\ln2\left(1+U_k[t]P_k[t]/\sigma_{\text{B}}^2\right)\log_2\left(1+U_k[t]P_k[t]/\sigma_{\text{B}}^2\right)\right]}{P_k^2[t]\ln2\left(1+U_k[t]P_k[t]/\sigma_{\text{B}}^2\right)}
	\end{equation}
	\hrulefill 
\end{figure*}
\begin{equation}
	\label{eqa3}
	\frac{\partial\eta_{\text{C}}}{\partial P_k[t]}=-\frac{2}{3}\cdot\frac{{\left( {1 - {\kappa _k}} \right)}}{{\gamma _c^{1/3}\left( {{e^{\varepsilon /{\kappa _k}}} - {e^\varepsilon }} \right)P_k^{5/3}[t]}}.
\end{equation}
$\frac{\partial\eta_{\text{U}}}{\partial P_k[t]} < 0$ when $P_k[t]\rightarrow0^+$. $\frac{\partial\eta_{\text{U}}}{\partial P_k[t]} < 0$ when $P_k[t]>0$. Therefore, the first-order partial derivatives of $\eta_{\text{U}}$ and $\eta_{\text{C}}$ are all less than 0 when $ P_k[t] >0$. $\eta_{\text{U}}$ and $\eta_{\text{C}}$ are monotonically decreasing when $P_k[t] >0$. Furthermore, we check the value of the energy efficiencies at the end of the define fields, i.e., at 0 and $+\infty$. We have $\lim\limits_{P_k[t]\rightarrow0^+}\frac{\eta_{\text{C}}}{\eta_{\text{U}}}\rightarrow+\infty$, i.e., $\eta_{\text{C}}>\eta_{\text{U}}$, and $\lim\limits_{P_k[t]\rightarrow+\infty}\frac{\eta_{\text{U}}}{\eta_{\text{C}}}\rightarrow+\infty$, i.e., $\eta_{\text{U}}>\eta_{\text{C}}$. Therefore, there must have exactly one $P^*$ that makes $\eta_{\text{U}}<\eta_{\rm{C}}$ when $P_k[t]<P^*$, $\eta_{\text{U}}\ge\eta_{\rm{C}}$ when $P_k[t]\ge P^*$. This completes the proof of Proposition 3.

\bibliographystyle{IEEEtran}
\balance
\bibliography{reference_MCS_UAV_RIS}

\end{document}